\documentclass[superscriptaddress,twocolumn, showpacs,amsmath,amssymb,aps,prl,floatfix, 10pt]{revtex4-1}
\usepackage[latin1]{inputenc}
\usepackage[american]{babel}
\usepackage[T1]{fontenc}
\usepackage{mathrsfs}
\usepackage{hyperref}
\usepackage{multirow}
\usepackage{bm, epsfig}
\usepackage{graphicx}
\usepackage{subeqnarray}
\usepackage{bbold}
\usepackage{upgreek}
\usepackage{tikz}
\usepackage{slashed}
\usepackage{lineno}
\hypersetup{colorlinks=true, urlcolor=blue, citecolor=blue, filecolor=blue, linkcolor=blue}

\newcommand{\eps}{\varepsilon}
\newcommand{\balpha}{\mbox{\boldmath $\alpha$} }

\usepackage{changes}
\definechangesauthor[name={Natalia}, color=magenta]{nat}

\usetikzlibrary{decorations.pathmorphing}
\usetikzlibrary{shapes}
\usetikzlibrary{shapes.geometric}

\tikzset{
    magnetic/.style={
        fill,
        shape border rotate=0,
        isosceles triangle,
        isosceles triangle apex angle=60,
        node distance=1,
        minimum height=.1
    }
}

\tikzset{
    othermagnetic/.style={
        fill,
        shape border rotate=90,
        isosceles triangle,
        isosceles triangle apex angle=60,
        node distance=1,
        minimum height=.1
    }
}

\begin{document}

\title{Self-energy correction to the energy levels of heavy muonic atoms}

\date{\today}

\author{Natalia~S.~Oreshkina}
\email[Email: ]{Natalia.Oreshkina@mpi-hd.mpg.de} 
\affiliation{Max-Planck-Institut f\"{u}r Kernphysik, Saupfercheckweg 1, 69117 Heidelberg, Germany}

\begin{abstract}
The first fully relativistic, rigorous QED calculations of the self-energy correction to the fine-structure levels of heavy muonic atoms are reported. 
We discuss nuclear model and parameter dependence for this contribution as well as numerical convergence issues.
The presented results show sizable disagreement with previously reported estimations, including ones used for the determination of the nuclear root-mean-square radii, and underline the importance of rigorous QED calculations for the theoretical prediction of the spectra of muonic atoms.
\end{abstract}


\maketitle




{\it Introduction.}
Muonic atoms are the bound systems of an atomic nucleus and a negatively charged muon. 
Being more than 200 times heavier than an electron, a muon possesses correspondingly downscaled atomic orbitals radii, which in the case of heavy muonic atoms are comparable or even smaller than the nuclear radius.
This leads to huge finite-nuclear-size effects and a strong dependence of the muon's bound-state energies on the nuclear charge and current distributions, as well as to large relativistic effects. 
The understanding of this strong dependence of the muonic atoms on nuclear parameters, and the information about atomic nuclei that they can deliver, has triggered interest in precise knowledge of the level structure of muonic atoms \cite{Wheeler1949, BorieRinker1982, Devons1995}. 
%
A combination of state-of-the-art theoretical predictions of the level structure and experiments measuring the transition energies in muonic atoms enabled the determination of nuclear parameters like nuclear-charge, or root-mean-square (rms) radii \cite{Pohl2010, Piller1990, Schaller1980, Saito2022}, electric quadrupole and magnetic dipole moments \cite{Antognini2020, Dey1979, Ruetschi1984, tanaka1983}. 
One of the most precise measurements to date is the determination of the rms radius of ${}^{208}$Pb to a 0.02\% level \cite{Bergem1988}.


The short life-time of muon leads to the fact that muonic atoms are essentially muonic hydrogenlike and can be described with the single-particle Dirac equation. 
The theory of muonic atoms, including nuclear and leading quantum electrodynamics (QED) corrections, has been presented already in Refs.~\cite{BorieRinker1982, wu1969}.
Recently the updated state-of-the-art calculations of the fine and hyperfine structures of heavy muonic atoms and the corresponding analysis of the individual contributions has been presented in Refs.~\cite{michel2017, Patoary2018, Michel2019ho, Indelicato2021, Okumura_2021}.
One of the important effects is the self-energy (SE) correction. 
Unlike the case of atomic electrons, where SE is comparable to another QED correction, the vacuum polarization (VP) correction \cite{Beier00}, in muonic atoms the VP correction is by far the dominant one 
\cite{BorieRinker1982}. 
Therefore, the SE correction is much smaller than the leading VP correction and was previously calculated 
within a relatively simple mean-value evaluation method, suggested in Ref.~\cite{Barrett1968} and later used in Refs.~\cite{BorieRinker1982, Haga2007}. 
Later, an attempt at a more precise calculation was  made in Ref.~\cite{Cheng1978} for the ground $1s_{1/2}$ state of several muonic atoms with a final uncertainty of about 5\%.
However, even most recent works on muonic atoms still exclude the SE correction from the theoretical description, and treat the leading VP correction as total QED contribution \cite{Indelicato2021, Saito2022}. 

Additionally, in some cases the analysis of high-precision spectroscopic x-ray measurements of the muonic transitions revealed some anomalies and disagreements with theoretical predictions.
Thus, the assumption about the most complicated nuclear polarization (NP) correction deduced from the experimental data for fine-structure components difference $\Delta 2p=E_{2p_{3/2}}-E_{2p_{1/2}}$ had an opposite sign compared to the theory results for ${}^{90}$Zr \cite{Phan1985}, ${}^{112-124}$Sn \cite{Piller1990}, and ${}^{208}$Pb \cite{Yamazaki1979, Bergem1988}. 
For a long time, it was believed that this anomaly can be explained by more precise predictions on the NP correction, but recently this was shown not to be the case \cite{Valuev2022}, and, therefore, additional attention should be paid to other contributions, in particular to the last remaining sizable QED effect, namely, to the SE correction.

In this Letter, we present rigorous, fully relativistic  QED calculations of the SE correction to the ground $1s_{1/2}$ and excited $2p_{1/2}$ and $2p_{3/2}$ state energies of muonic atoms,
and establish the accuracy of our predictions for several muonic atoms of interest.
The results can be used for future experiments aiming at high-precision determination of nuclear rms radii, and for reanalyzing the existing experimental data in order to resolve the fine-structure anomaly.


{\it Formalism.}
The method for calculation of the SE correction for the bound electron in highly-charged hydrogen-like ions was first proposed in Ref.~\cite{Brown_1959}, used in Ref.~\cite{Desiderio_1971}, and further improved in Refs.~\cite{Mohr_1982, Indelicato_1992}. 
In our current work, we apply the procedure described in detail in \cite{Yerokhin_1999, Oreshkina2018_sewf}
with an emphasis on the finite-nuclear size effects.

The self-energy correction to the state $a$ with energy $\varepsilon_a$ can be written in terms of matrix element of $\Sigma(E)$ in the Feynman gauge as \cite{Yerokhin_1999, Oreshkina2018_sewf}:
\begin{align}\label{eq:SEstart}
&\langle a|\Sigma(\varepsilon_a)|a\rangle = \frac{i}{2\pi}\int_{-\infty}^{\infty} {\rm d} \omega
\sum_n \frac{\langle an|I(\omega)|na\rangle}{\varepsilon_a-\omega-\varepsilon_n(1-i0)},
\end{align}
\begin{align}
&I(\omega,\mathbf{x}_1,\mathbf{x}_2) = \alpha \frac{(1-\balpha_1 \balpha_2)
\exp{(i\sqrt{\omega^2+i 0}x_{12})}}
{x_{12}}. 
\end{align}
Here, $x_{12}$ is the relative distance $x_{12} = |\mathbf{x}_1-\mathbf{x}_2|$, $\balpha$ are the Dirac matrices, $\omega$ is the energy of a virtual photon, and the summation 
goes over the full spectrum $n$ 
of the considered lepton (electron or muon), including negative- and positive-energy states. 
The muonic relativistic system of units ($\hbar = m_\mu = c = 1$) and Heaviside charge units ($\alpha = e^2/(4\pi)$) are used throughout the paper.  
Bold letters are used for 3-vectors, the components of 3-vectors are listed with Latin indices, whereas  Greek letters denote 4-vector indices.
The usage of the relativistic muonic system of units with $m_\mu=1$ allows us to use exactly the same formulas which were derived for electronic atoms in the relativistic electronic system of units with $m_e=1$, and the difference appears only in the value of the rms radius.

Following the procedure and notations from Refs.~\cite{Yerokhin_1999, Oreshkina2018_sewf}, we expand the  Green's function for the bound electron in powers of the Coulomb potential. 
After this, taking into account the mass counterterm, and performing the angular integration analytically, one can write the total SE correction for a state $a$ as the sum of non-divergent zero-potential, one-potential and many-potential terms, respectively:
\begin{align}
\Delta E_{a}^{\rm{SE}} &= \Delta E_{a}^{(0)} + \Delta E_{a}^{(1)} + \Delta E_{a}^{(2+)}.
\end{align}
The zero-potential term is calculated in the momentum representation as the diagonal SE matrix element:   
\begin{align}\label{eq:0pot_final}
\Delta E_{a}^{(0)} & = \frac{\alpha}{4\pi} 
\int_0^{\infty} \frac{{\rm d}p \,p^2} {(2\pi)^3} \bigl\{
a(\rho)(g_a^2-f_a^2)  \\
&+ b(\rho) [\eps_a(g_a^2+f_a^2) +
2pg_af_a]\bigr\}, \notag 
\end{align}
where $g_a$ and $f_a$ are large and small components of the radial wave function of the state $a$ with the energy $\eps_a$ in the momentum representation, $\rho = 1 - {\rm p}^2((e^2+p^2))$, and the functions $a(\rho)$ and $b(\rho)$ are given in Ref.~\cite{Oreshkina2018_sewf}.

The one-potential term has one single interaction between the electron and the nucleus inside the SE loop, and its final renormalized expression can be written in the momentum representation as \cite{Snyderman_1991,Yerokhin_1999, Oreshkina2018_sewf}

\begin{align}\label{eq:1pot_final}
\Delta E_{a}^{(1)} &= \frac{\alpha}{2(2\pi)^6} 
\int_0^{\infty} {{\rm d}p\,p^2} \int_0^{\infty} {{\rm d}p'\,p'^2}  \\ 
& \times\int_{-1}^{1}{\rm d}\xi \, V(q)[{\cal F}^{aa}_1P_{l_a}(\xi) + {\cal F}^{aa}_2P_{\overline{l}_a}(\xi)], \notag
\end{align}
where $q^2 = p^2 +p'^2 - 2pp'\xi$, $\overline{l}$ is defined through the total angular momentum $j$ and orbital angular momentum $l$ as $\overline{l} = 2j-l$, $P_l$ are the Legendre polynomials, and ${\cal F}^{ab}_{1,2}$ contain one more internal integration and are given in Ref.~\cite{Oreshkina2018_sewf}.
The exact expressions for the nuclear potential $V(q)$ in the momentum representation will be discussed later.

Finally, the many-potential term with two or more interactions between the electron and the nucleus inside the SE loop can be calculated in the coordinate representation, and performing  the angular integrations and summations analytically one gets:
\begin{align}\label{eq:2pot_final}
\Delta E_{a}^{(2+)} &= \frac{i\alpha}{2\pi}
\int_{-\infty}^{\infty}{\rm d}\omega \sum_{nJ}
\frac{(-1)^{j_n-j_a+J}}{2j_a+1} \notag \\
&\times \frac{R_J(\omega, an'n'a)}
{\eps_a-\omega-\eps_n(1-i0)},
\end{align}
where $\omega$ corresponds to the energy of the virtual photon, $R_J(\omega, abcd)$ are generalized Slater radial integrals \cite{Oreshkina2018_sewf}, the sum over $n$ in this expression corresponds to the summation over the intermediate states, and 
\begin{align} \label{eq:irr2+}
|n'\rangle &= \sum_f \frac{|f\rangle\langle
f|V|n\rangle}{\eps_a-\omega-\eps_f(1-i0)}.
\end{align}
Here $|n\rangle$ is a bound state in the external nuclear potential~$V$, and $|f\rangle$ belongs to the spectrum of the free electron.

{\it Nuclear potentials in coordinate and momentum representations.}
In addition to the wavefunction in the momentum representation for the calculation of zero- and one-potential terms \eqref{eq:0pot_final}-\eqref{eq:1pot_final}, one-potential contribution also contains the nuclear potential.
Calculated with a generalized Fourier transform \cite{Oreshkina2018_sewf}, 
the Coulomb potential for point-like nucleus 
\begin{align}
V_{\rm Coul}(r) = -\frac{\alpha Z}{r},
\end{align}
has the following view in the momentum representation:
\begin{align}
V_{\rm Coul}(q)= -4\pi\frac{\alpha Z}{q^2}.
\end{align}
However, the Coulomb potential should not be applied in the case of muonic atoms due to the significant nuclear effects, and therefore in the current work it has been replaced with different finite-nuclear-size potentials.
The first of them, the shell distribution model, has a simple analytical form in both coordinate and momentum representations:
 \begin{align}
V_{\rm shell}(r) & = 
\begin{cases} 
    -\cfrac{\alpha Z}{R_0} & r< R_0 \\
    -\cfrac{\alpha Z}{r} & r>R_0
    \end{cases}, \\
\label{eq:sh_mom}
V_{\rm shell}(q) & = -4\pi\frac{\alpha Z}{q^2} \frac{\sin{R_0q}}{R_0q}.
\end{align}
Here, the parameter $R_0$ is defined in terms of the rms radius as $R_0 = \sqrt{\langle r^2 \rangle}$.
A nuclear model which assumes homogeneous-sphere charge distribution of the charge density corresponds to the following potential in the coordinate and momentum representations:
\begin{align}
V_{\rm sph}(r) &= 
\begin{cases} 
    -\cfrac{\alpha Z}{R_0} \biggl[\cfrac{3}{2} - \cfrac{1}{2} \biggl(\cfrac{r}{R_0}\biggr)^2 \biggr] & r< R_0 \\
    -\cfrac{\alpha Z}{r} & r>R_0
    \end{cases}, \\
\label{eq:sph_mom}
V_{\rm sph}(q) &= -4\pi\frac{\alpha Z}{q^2} \frac{3(\sin{R_0q} - R_0q \cos{R_0q})}{(R_0q)^3}.
\end{align}
The parameter $R_o$ is now defined as $R_0 = \sqrt{5/3 \langle r^2 \rangle}$. 
Finally, we also used the most realistic Fermi distribution of the nuclear density \cite{Parpia1992}:
\begin{align}
\rho = \frac{\rho_0}{1+e^{(r-c)/a}}.
\end{align}
Here, $a$ is the skin thickness and it is usually assumed to be $a=2.3{\text{ fm}}/4\log(3)$ \cite{Parpia1992,Beier00}.
The condition that $V(r)$ has to be normalized to the nuclear charge $Z$ defines normalization constant $\rho_0$, and
the half-density radius $c$ is chosen to reproduce the rms value.
The analytical formula for the nuclear potential created by Fermi nuclear charge distribution is then \cite{Parpia1992}:
\begin{widetext}
\begin{subequations}
\begin{align}
V_{\rm Fermi}(r<c) & = -\frac{\alpha Z}{r} \frac{1}{N_{\rm Fermi}} \biggl\{ 
	6 \biggl( \frac{a}{c}\biggr)^3 \biggl[ S_3\biggl( \frac{r-c}{a}\biggr) - S_3\biggl( -\frac{c}{a}\biggr) \biggr]
	+ \frac{r}{c}\biggl[ \frac{3}{2} + \frac{\pi^2}{2}\biggl( \frac{a}{c}\biggr)^2  
	- 3 \biggl( \frac{a}{c}\biggr)^2 S_2\biggl( \frac{r-c}{a}\biggr) \biggr]
	- \frac{1}{2} \biggl( \frac{r}{c}\biggr)^3 \biggr\}, \\
V_{\rm Fermi}(r>c) & = -\frac{\alpha Z}{r} \frac{1}{N_{\rm Fermi}} \biggl\{ 
	N_{\rm Fermi} + 6 \biggl( \frac{a}{c}\biggr)^3 S_3\biggl( \frac{c-r}{a}\biggr) + 
	3 \biggl( \frac{a}{c}\biggr)^2 \frac{r}{c} S_2\biggl( \frac{c-r}{a}\biggr)\biggr\}, \\
{\text{where}} \qquad S_k(x) & = \sum_{n=1}^\infty \frac{(-1)^n}{k^n} \exp(nx) \qquad \text{and} \qquad
N_{\rm Fermi} = 1 + \pi^2 \biggl( \frac{a}{c}\biggr)^2 -  6 \biggl( \frac{a}{c}\biggr)^3 S_3\biggl( -\frac{c}{a}\biggr).
\end{align}
\end{subequations}
After performing a Fourier transform and calculating the integral analytically, we get:

\begin{subequations}\label{eq:Fermi_mom}
\begin{align}
V_{\rm Fermi}(q) & =  V_{\rm Coul}(q) \biggl( 1 +  \frac{\widetilde{V}_{\rm triv}(q)}{N_{\rm Fermi}} 
		+ \frac{\widetilde{V}_{\rm sum}(q)}{N_{\rm Fermi}} \biggr) \\
\widetilde{V}_{\rm triv}(q) & = -\frac{c^2 + a^2\pi^2}{c^2} + \frac{1}{(cq)^2} 
	\biggl( \frac{a^2q^2\pi^2-6}{2} \cos (cq) + \frac{a^2q^2\pi^2+6}{2} \frac{\sin(cq)}{cq}\biggr) \\
\widetilde{V}_{\rm sum}(q) & = 6 \biggl( \frac{a}{c}\biggr)^2 \cos (cq) \biggl[ \frac{1}{2(aq)^2} + \frac{\pi^2}{6} 
		- \frac{\pi}{2(aq)} \coth(\pi aq) \biggr]	
		+ 6 \biggl( \frac{a}{c}\biggr)^2 \frac{\sin(cq)}{cq} \biggl[ -\frac{1}{2(aq)^2} + \frac{\pi^2}{6} 
		+ \frac{\pi^2}{2} \frac{1}{\sinh^2 (\pi aq)} \biggr] \notag \\
		& + 6 \biggl( \frac{a}{c}\biggr)^3 (aq)^2 \sum_{n=1}^{\infty} \frac{2n^2 + (aq)^2}{n^3[n^2 + (aq)^2]^2}
		\exp(-nc/a). \label{eq:Fermi_mom_c}
\end{align}
\end{subequations}
\end{widetext}
The only singular contribution in Eq.~\eqref{eq:Fermi_mom} coincides with the Coulomb part $V_{\rm Coul}(q)$; all remaining coefficients and contributions are regular at $q=0$ even though it can be not obvious from the expressions.   
The potential itself is given in terms of elementary functions and can be easily implemented for numerical calculation with the single exception of the last term in Eq.~\eqref{eq:Fermi_mom_c}, which nevertheless converges very fast and therefore does not limit the numerical accuracy.


{\it Calculation details.}
For the numerical integrations we used the numerical solution of the Dirac equation utilizing the dual-kinetic-basis (DKB) approach \cite{dkb} involving basis functions represented by piecewise polynomials on grid's intervals from B-splines. 
This method allows one to find solutions of the Dirac equation for an arbitrary spherically symmetric potential in a finite-size cavity, and describes both the discrete and continuous spectra with a finite number of electronic states for every given $j$ and $l$. 

For the numerical evaluation in Eqs.~\eqref{eq:0pot_final} and \eqref{eq:1pot_final}, routines from the numerical integration library QUADPACK \cite{quadpack} have been used for the generalized Fourier transformation of the wave functions and further integrations, following the methods developed in \cite{Oreshkina2018_sewf}.
Analytical formulas~\eqref{eq:sh_mom}, \eqref{eq:sph_mom} and \eqref{eq:Fermi_mom} for the nuclear potential in momentum representation have been used for the shell, sphere and Fermi models, correspondingly.  

\begin{figure*}[ht!]
\includegraphics[trim = 0cm 4.5cm 0cm 2cm, clip, width=0.8\paperwidth]{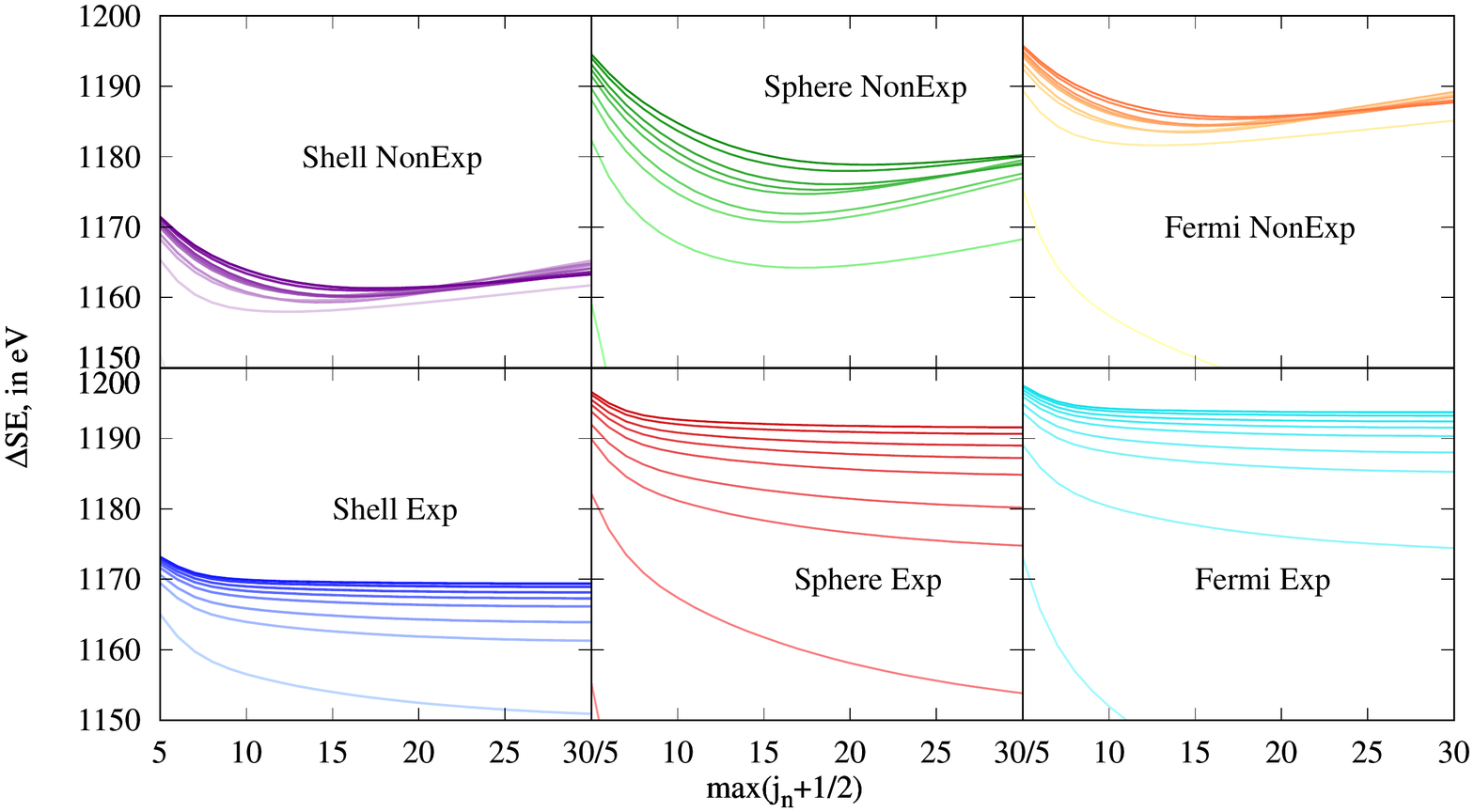}
\caption{$\Delta E_{\rm SE}$ contribution to the $1s_{1/2}$ state of the muonic zirconium in units of eV as a function of maximal intermediate angular momentum $j_n$ for different nuclear models and numerical grids. The colors of the lines on every panel change depending on the number of used DKB basis functions from light for $n_{\rm DKB}=50$ to dark for $n_{\rm DKB}=150$.
} \label{fig:40all}
\end{figure*}

\begin{figure}[ht!]
\includegraphics[width=0.40\paperwidth]{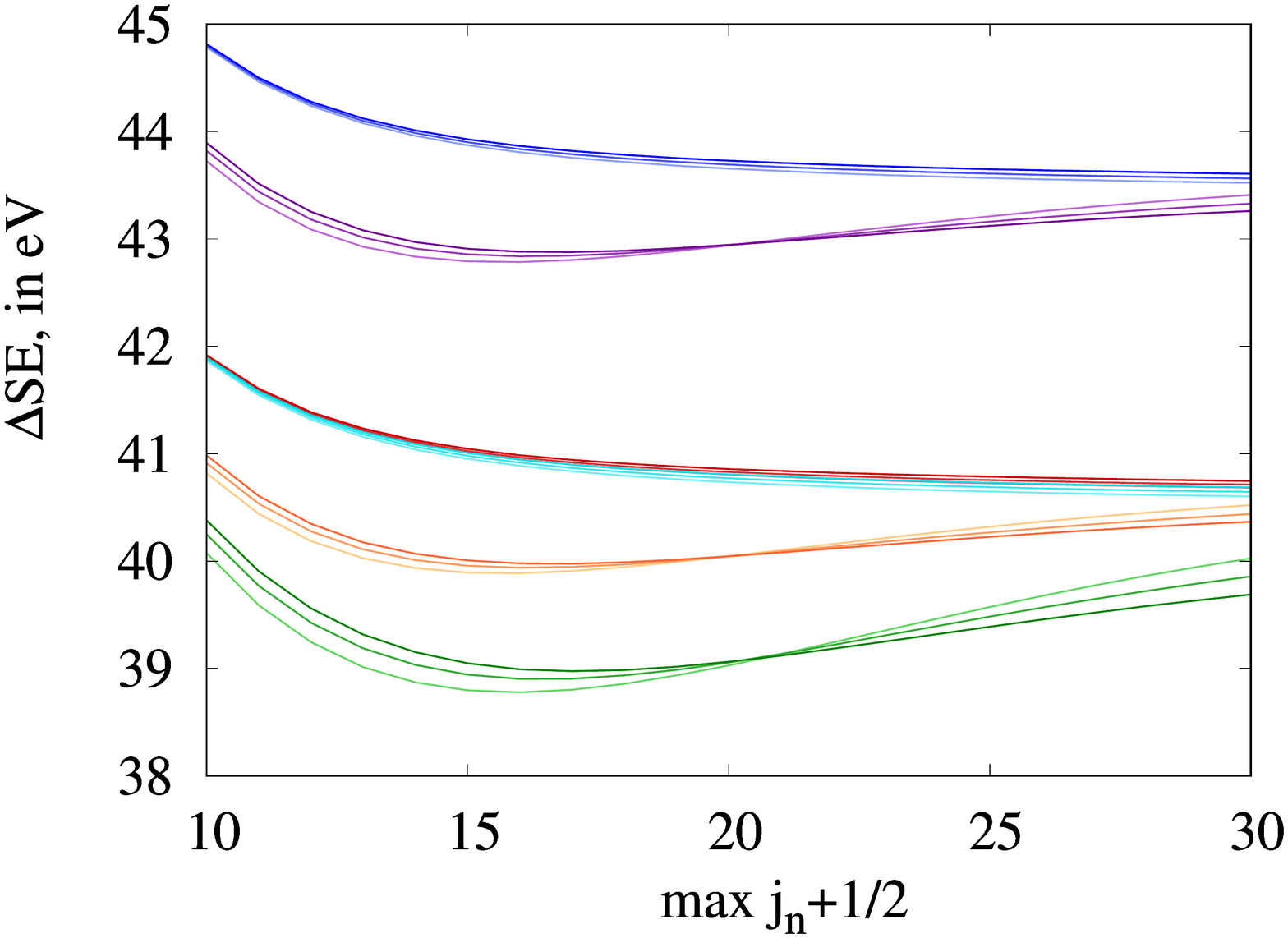}
\includegraphics[width=0.40\paperwidth]{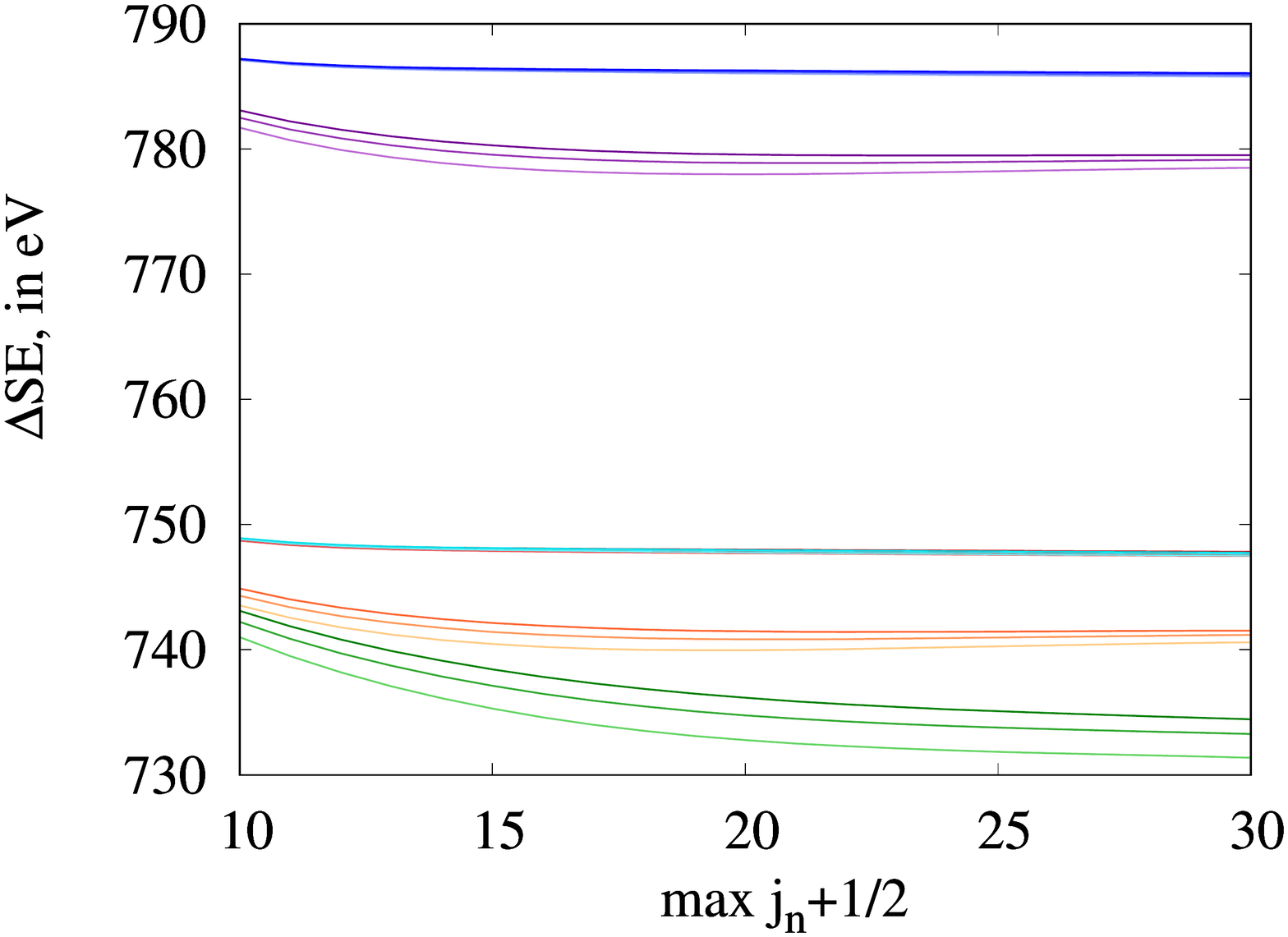}
\caption{$\Delta E_{\rm SE}$ contribution to the $2p_{1/2}$ state of muonic tin (top) and $2p_{3/2}$ state of the muonic lead (bottom) in units of eV as a function of maximal intermediate angular momentum $j_n$ for different nuclear models and numerical grids. The color scheme in respect to the nuclear models and numerical grids corresponds to the one used in Figure~\ref{fig:40all}, the colors of the lines change depending on the number of used DKB basis functions from light for $n_{\rm DKB}=130$ to dark for $n_{\rm DKB}=150$.
} \label{fig:40_2p3_all}
\end{figure}

The summation in the remaining many-potential term over the intermediate state $n$ in Eq.~\eqref{eq:2pot_final} goes over the principal quantum number and at the same time involves an infinite summation over the total angular momentum $j_n$. 
Ideally, the calculations with infinite number of basis functions and with the summation which extends up to $j_n=\infty$ would give the most accurate result,  but in reality reaching infinity in both of these directions is neither possible nor necessarily beneficial.
The DKB wavefunctions of low-lying bound states are reproduced with very high accuracy, and the summations over the Dirac spectrum can be performed very well.
However, for the states with high values of angular momentum $j$ even the lowest-lying states have large numbers of knots and oscillate, and therefore the accuracy of the calculations cannot be improved by a simple increase of the basis. Therefore, in our numerical calculations, the individual terms up to the maximum value $j_n+1/2 = 30$ have been calculated to analyze the convergence.
Additionally, for every nuclear model (shell, sphere and Fermi) and for two different types of the DKB grid (exponential and non-exponential) it has been performed for the number of basic functions $n_{\rm DKB}$ increasing from 50 up to 150.

The corresponding results for the $1s_{1/2}$ state of muonic zirconium are presented in Figure~\ref{fig:40all}. 
The colors of the lines on every panel change from light to dark for lower to higher number of basis functions.
As one can see from the Figure, the convergence is better for the exponential grid since the results are stable with respect to the maximal $j_n$, whereas for the non-exponential grid, even though the lines corresponding to the different $n_{\rm DKB}$ are closer to each other, they still change visibly as functions of~$j_n$. 
The calculated SE correction is rather sensitive to the maximum value $j_n$, the number of used basic functions $n_{\rm DKB}$, the nuclear model and the integration grid, so only a combined deep analysis of these dependencies would allow us to give a reliable and accurate prediction of the effect.  



{\it Results.}
In our current work, we focus on the low-lying states with high importance for experimental analysis for the muonic atoms whose spectra have already been measured before.
The total value of the  SE correction for the $1s_{1/2}$, $2p_{1/2}$ and $2p_{3/2}$ states of muonic zirconium, tin and lead for three nuclear models and two DKB grids are listed in Table \ref{tab:01}, for $\max(j_n+1/2)=30$ and $n_{\rm DKB}=150$. 
To be conservative, we take an average of two grids for the Fermi model as the final one, and estimate the uncertainty by the
comparison between different models and grid predictions; however, as one can see from our results, for heavy nuclei the shell model results deviate more and more from those of the Fermi and sphere models, confirming the low applicability of the this model for heavy muonic atoms.
We have also estimated the dependence of our results from the used rms value, and even in the most sensitive case of the $1s_{1/2}$ state it is on the level of 0.1\% and well below the nuclear model dependence.
The comparison with the current state-of-the-art theoretical predictions from Refs.~\cite{Haga2007, Cheng1978} shows disagreement, sometimes even outside of the few-percent uncertainties indicated there, and an order-of-magnitude improvement in the accuracy.
However, despite the disagreement in individual contributions, the fine-structure component difference $\Delta 2p$ is in  surprisingly good agreement with the earlier predictions, and therefore one can conclude that the last sizeable QED effect, namely the SE correction, does not resolve fine-structure muonic anomaly. 
Finally, the updated rigorous SE results to the transition energies can potentially change the rms values based on the muonic spectra and play important role in future for new experiments aiming to the extraction of nuclear moments and rms radii.


\begin{table*}
\begin{center}
\begin{tabular}{c l c c c c c c c c c c }
\hline 
\hline 
Ion & State	& Shell Exp & Shell NonExp & Sphere Exp & Sphere NonExp & Fermi Exp & Fermi NonExp & Final & Previous \\ 
\hline 
$\mu-{}^{90}$Zr & $1s_{1/2}$ & 1169.4 & 1163.3	& 1191.6 & 1180.2 & 1193.7	& 1187.7 & 1191(4) & 1218 \\
				& $2p_{1/2}$ & 7.675  & 7.601	& 7.055  & 6.987  & 7.031	& 6.963  & 6.99(5) & 1 \\
				& $2p_{3/2}$ & 47.16  & 47.05	& 46.59  & 46.55  & 46.58	& 46.47	 & 46.52(6) & 41 \\
\hline
$\mu-{}^{120}$Sn & $1s_{1/2}$ & 1677.6	& 1665.4 & 1725.9 & 1701.0	& 1729.7 & 1717.7	& 1724(7) & --- \\
				& $2p_{1/2}$  & 43.61	& 43.26	 & 40.75  & 39.86	& 40.69	 & 40.37	& 40.5(3) & --- \\
				& $2p_{3/2}$  & 123.1	& 122.7	 & 120.5  & 119.8	& 120.5	 & 120.1	& 120.3(3) & --- \\
\hline
$\mu-{}^{208}$Pb & $1s_{1/2}$ & 3041.4	& 3012.0 & 3229.4 & 3197.4	& 3239.4 & 3210.8	& 3225(15) & 3373 \\
				& & & & & & & & & 3270(160)~\cite{Cheng1978}\\
				& $2p_{1/2}$  & 497.6	& 490.6	& 457.1	& 440.9		& 456.7	& 450.0		& 453(5) & 413 \\
				& $2p_{3/2}$  & 786.0	& 779.5	& 747.8	& 734.4		& 747.8	& 741.5		& 745(5) & 707 \\
\hline
\hline 
\end{tabular}
\caption{$\Delta E_{\rm SE}$ contribution to the low-lying states to bound-muon energy in units of eV. 
One but last column corresponds to our final value with an errorbar.
The previous results have been taken from Ref.~\cite{Haga2007}, and the one taken from  Ref.~\cite{Cheng1978} is noted  explicit.
} 
\label{tab:01}
\end{center}
\end{table*}

{\it Conclusions.}
We presented the first rigorous calculation of the SE correction to the few first energy levels of muonic atoms. 
We used three different nuclear charge distributions and two grids to estimate the convergence and uncertainties of our predictions, and gave an analytical formula for the Fermi potential in momentum representation.
Theoretical results for low-lying $1s_{1/2}$, $2p_{1/2}$ and $2p_{3/2}$ for muonic zirconium, tin and lead have been presented and compared with the available published results.
This comparison shows significant difference between our rigorous calculations and the previous mean-value method, therefore justifying the usage of the accurate QED approach for high-precision calculations.
Even though the renewed SE results do not resolve the fine-structure muonic anomaly, they can affect previous results of the extraction of the nuclear rms radii based on the muonic atoms spectra, and stimulate the search of other sources for this problem, including physics beyond the Standard Model.
More importantly, the rigorous high-precision SE results are an essential part of the state-of-the-art theoretical predictions aiming at a high-precision determination of nuclear radii and other parameters. 

The Author thanks I.~A.~Valuev, Z.~Harman, V.~A.~Yerokhin, and D.~A.~Glazov for useful discussions.

\end{document}